\documentclass[aip,rsi,reprint,graphicx]{revtex4-1} % for checking your page length
\usepackage{graphicx}
\usepackage{xcolor}

\begin{document}

%\preprint{}

\title{Miniature Dilatometer Based upon an AFM Piezocantilever} %Title of paper



\author{L. Wang}\email[]{wangliran@magnet.fsu.edu}
\affiliation{National High Magnetic Field Laboratory, Florida State University, Tallahassee, Florida 32310, USA}
\author{G. M. Schmiedeshoff}
\affiliation{Department of Physics, Occidental College, Los Angeles, California 90041, USA}
\author{D. Graf}
\affiliation{National High Magnetic Field Laboratory, Florida State University, Tallahassee, Florida 32310, USA}
\author{J. -H. Park}
\affiliation{National High Magnetic Field Laboratory, Florida State University, Tallahassee, Florida 32310, USA}
\author{T. P. Murphy}
\affiliation{National High Magnetic Field Laboratory, Florida State University, Tallahassee, Florida 32310, USA}
\author{S. W. Tozer}
\affiliation{National High Magnetic Field Laboratory, Florida State University, Tallahassee, Florida 32310, USA}
\author{J. L. Sarrao}
\affiliation{Los Alamos National Laboratory, MST-10, Los Alamos, New Mexico 87545, USA}
\author{J. C. Cooley}
\affiliation{Los Alamos National Laboratory, MST-10, Los Alamos, New Mexico 87545, USA}
\author{E. C. Palm}
\affiliation{National High Magnetic Field Laboratory, Florida State University, Tallahassee, Florida 32310, USA}
%\thanks{}


\date{\today}

\begin{abstract}
 We report on the development of a sensitive dilatometer based upon a AFM piezocantilever.  This dilatometer has been tested at temperatures down to 25 mK and in magnetic fields up to 16 T. The layered heavy fermion superconductor $CeCoIn_5$ and its non-magnetic analog $LaRhIn_5$ have been measured to demonstrate its use in detecting phase transitions and quantum oscillations. In addition, using this dilatometer, a simultaneous multi-axis dilation measurement has been done. This compact dilatometer has many advantages such as its ability to measure very small samples with lengths at sub-mm levels, low temperature and field dependence, ability to rotate, and works well irrespective of being in a changing liquid or gas environment (i.e. within a flow cryostat or mixing chamber).
\end{abstract}


\maketitle %\maketitle must follow title, authors, abstract and \pacs

% References should be done using the \cite and \label commands
\section{Introduction}\label{introduction}

Thermal expansion (TE) and magnetostricton (MS) can provide accurate information on 
changes in the dimensions of
materials as the temperature and magnetic field are varied. These quantities are of considerable
interest because of the fundamental importance of the structure of materials
and their intimate relation to the specific heat of materials\cite{Barron1999,Grueneisen}. 
They provide a way to understand the pressure dependencies of the respective ordering phenomena, thermodynamic properties, relevant energy changes, and magnetic field-induced structural changes,
which are very important for the study of high temperature superconductors
\cite{Lortz2003,Meingast2001}, heavy fermion
systems\cite{Schmiedeshoff2011,AdeVisser1990} and many other novel materials. Unfortunately, many samples of interest are frequently small with a correspondingly small TE or MS, so a dilatometer with a high resolution is needed for these measurements.

Currently, the dilatometers based upon the capacitance method are one of the most sensitive
methods for precise TE and MS measurements\cite{White1993,White1974,Green1963,Pott1983,Swenson1998,Rotter1998}. Dilatometers of this design have operated successfully in a wide variety of different cryostats and measurement systems.  Their main advantages are that the sample is only under very weak uniaxial stress (50-500 mN)\citep{Rotter1998} and that a dilation limit of less than 0.1$\AA$ can be reached\citep{Schmiedeshoff2006}. However, when placed into a low temperature environment that changes from liquid to gas (as inside a He-3 cryostat) or even in a dilution refrigerator mixing chamber where the ratio of $^3$He to $^4$He changes with temperature or field, the capacitance dilatometer has large background  effects due to the temperature dependent dielectric constant of the medium. In addition, in a pulsed magnetic field, where the field is changing extremely rapidly, eddy currents in the metallic body of the device can have deleterious effects.  Finally, many capacitance dilatometers will change their capacitance upon rotation of the device due to the effect of gravity even without the sample length having changed, making their use in rotators difficult\cite{Brown1983,Johansen1986,Neumeier2008}.

To solve these problems, several solutions have been proposed. One is an optical fiber strain gauge using Fiber Bragg Gratings (FBGs). This method has the significant advantage of immunity to electromagnetic interference and works very well in a wide variety of cryostats and large pulsed magnetic fields\cite{2010RScI81c3909D}. However, the FBG has to be glued to the surface of the sample and this requires that the sample has a flat surface at least a few millimeters in length which is not possible with many samples. In addition, it is difficult to rotate the device and bend the optical fiber.

The solution proposed here uses an atomic force microscope piezo-resistive cantilever as the sensing element. The prototype construction was described earlier\cite{Park2009} and is here developed into a more robust device. A significant advantage of this device is its immunity to the effects of the cryogenic media it is immersed in. In addition, the cantilever is small and easy to mount in a relatively small space and can be easily rotated. Another advantage is the ability to perform simultaneous mulit-axis measurements. For previous dilatometer forms (capacitive and optical), dilations of the sample can only be measured one axis at a time. With proper construction, more than one cantilever can be used at the same time and measured separately  in order to detect the simultaneous changes of different axes of one sample; $x$, $y$ and even $z$ axis.

This work concentrates on the following aspects: finding suitable construction materials and design, exploring low noise and high-quality measurements, and discovering applications and limitations for this method.

\section{Basic measurement set-up}

\begin{figure}
\includegraphics[width=1\linewidth]{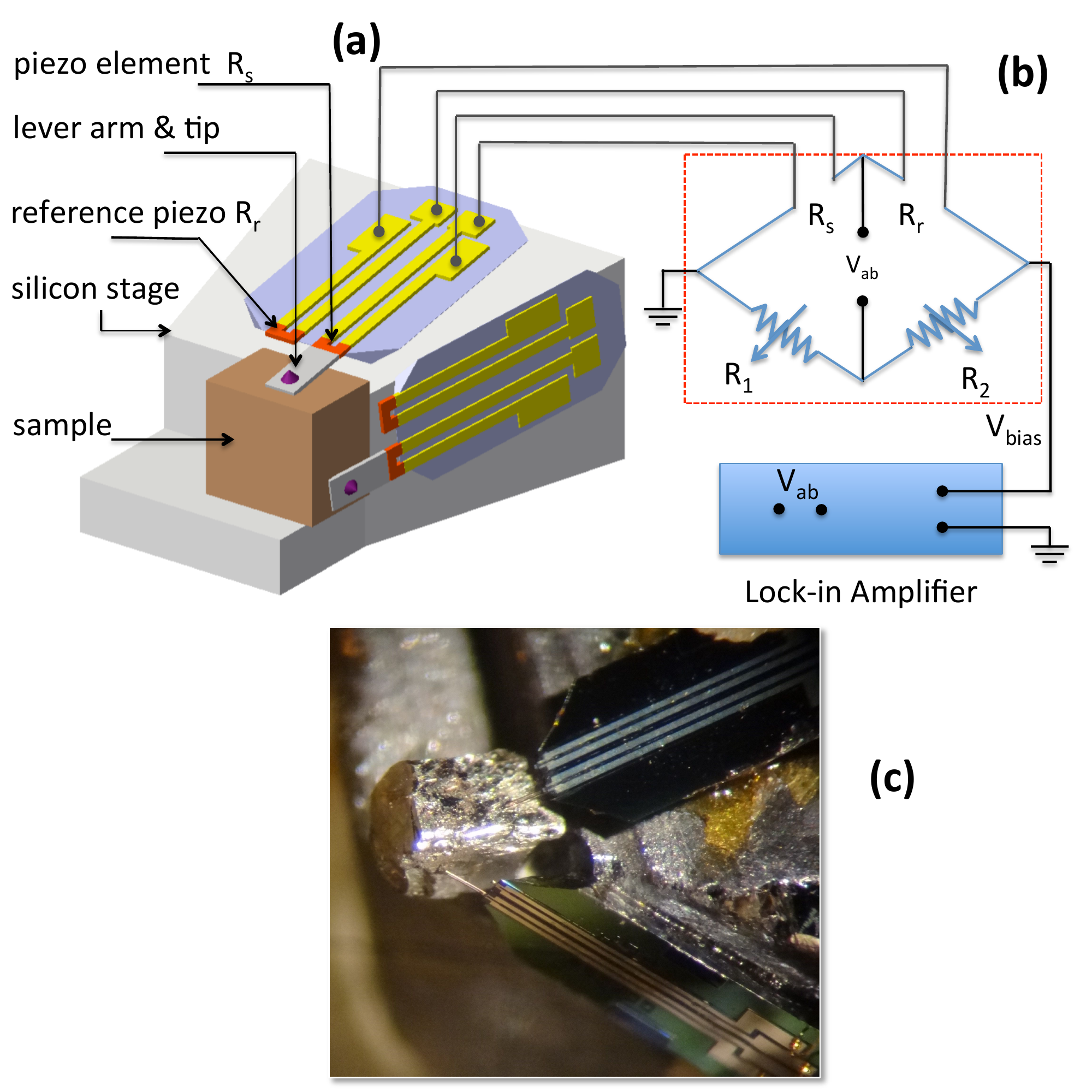}% % Important NOTE: Please make certain your figures do not include local directory paths. ex."c:\file\sub\fig1.eps" 
\caption[] {\label{1} Construction for a 2-axis measurement. (a) The silicon stage with two cantilever tips touching the sample surface. The cantilever arm and tip, piezo element $R_s$ and reference piezo $R_r$ are indicated. (b) The construction the Wheatstone bridge circuit. $V_{bias}$ is the output voltage, and $V_{ab}$ is the recorded signal. The circuit within the red box has been synthesized by a low noise circuit bridge, designed and built by the NHMFL electronic shop. (c) Image of the measurement setup as two cantilever tips touch the two surfaces of $CeCoIn_5$ separately as illustrated by (a).} \end{figure}

The dimensions of the commercial PRC400 \citep{Seiko} device measure
3.5$\times$1.6$\times$0.2 mm$^3$ (L$\times$W$\times$H) overall and the sample lever arm has
dimensions of 0.4$\times$0.05$\times$0.005 mm$^3$ (L$\times$W$\times$H). $R_{s}$
indicates the resistance of piezoelement of the long-tip lever arm and $R_{r}$ is the resistance of a short-tip
reference lever which is synthesized together with the signal piezo element. The typical
room temperature resistance of each piezo (both $R_{s}$ and $R_{r}$) is about 600-700 $\Omega$. As shown in Fig.\ref{1} (b),  the basic measurement circuit of  is  based on a Wheatstone bridge configuration. Resistance changes in the piezoelement $R_s$ that are induced by dimensional changes from the sample are recorded as a voltage signal in a lock-in amplifier. $V_{bias}$ is the input excitation voltage and $V_{ab}$ is the recorded output signal. $R_{1}$ and $R_{2}$ are the potentiometers used to balance the bridge. The equation is:
\begin{equation}
V_{ab}=V_{bias}(\frac{R_{2}}{R_{s}-R_{2}}-\frac{R_{1}}{R_{r}-R_{1}})
\end{equation} 

A top-loading $^3$He/$^4$He dilution refrigerator and superconducting magnet were used to provide temperatures as low as $\sim$25 mK and magnetic fields up to 16 T. The dilatometer is immersed directly in the $^3$He-$^4$He mixture providing excellent thermal contact to  the sample. 

For designing this dilatometer, there are two basic rules: First, the design must be very simple,  since any strain in the construction can transfer to the output signal, and the sample signal will normally be hidden by this huge effect. Second, it should be made of the same material as the cantilever body. If the materials are different, the differential contraction between the arm and the base can cause the cantilever tip to "walk" over the surface of the sample and produce erroneous signals. The final design of sample stage is made of silicon and the height of the stage where the sample sits is machined according to the thickness of the sample. The cantilever and sample are both glued to the silicon base with superglue and the sensory tip of the cantilever gently touches the sample surface. This simple design is especially suitable for thin samples with very limited length, $L_{sample}$ $\sim$ 0.1 mm or smaller. For a bulk sample with larger size, by using multiple cantilevers, the dilations of more than one sample direction can be measured simultaneously. The 2-axis measurement setup is shown as an example in Fig.\ref{1} (a) and with proper construction, measuring a 3$^{rd}$ axis is also possible.

Fig.\ref{1} (b) shows the Wheatstone bridge circuit construction. This part is synthesized in a custom manufactured low noise breakout box. This approach dramatically reduced the connections and exposed wires that can introduce unwanted noise. Signal Recovery (SR) 7280 lock-in amplifiers are used here. Typical settings were $V_{bias}$ = 0.05 V  with time constant 500 ms and excitation frequency f = 17 Hz. With the improved circuit, the peak-to-peak noise level can be reduced down to 20 nV.

In capacitance dilation measurements, $L$ represents the sample length and $\Delta L$ is used to quantify the change. In the dilatometer measurements described here, results are recorded in the form of electronic signal, Voltage. The signal $V_{ab}$ can reflect the deviation of the cantilever tip $\Delta L$. At room temperature, the calibration from $V_{ab}$ to $\Delta L$ can be made by using a micrometer to push the cantilever tip (record $\Delta L$) and recording the electrical signal ($V_{ab}$) at the same time. However the calibration relation is not as easily obtained for low temperature environments. There is an important factor which will dramatically influence the calibration, that can best be described in terms of the ``cell effect".

In a capacitive dilatometer, the cell effect is caused by the thermal expansion and magnetostriciton of the material(s) making up the dilatometer. In the cantilever dilatometer, the cell effect originates from two piezoresistors that do not match. The two resistors were presumably engineered to be identical at room temperature and in zero fields. What has been observed, though, is that small differences between them increase at low temperatures (slightly different temperature-dependent resistivities) and high fields (slightly different magnetoresistances). To create a new term, this might be called a ``resistor mismatch cell effect". Based on the tests of several cantilevers, this kind of cell effect can clearly vary from cantilever to cantilever. So for low temperature calibration, the conversion form $V_{ab}$ to $\Delta L$ cannot be a universal value for all cantilevers.

In spite of this, for repeated measurements with one single cantilever, the cell effect is rather reproducible. Most of the time, the cell effect is shown as a very broad bump or nearly linear background signal throughout the whole measurement range. Most of the dilation changes that are associated with phase transitions can be easily resolved as anomalies in the broad background signal. A rough estimation was made measuring one sample with both capacitance and piezo dilatometers. Under the same conditions (pressure, temperature and magnetic field), the dilation change appeared to be comparable. Here we use the same $CeCoIn_5$ sample which is used in Correa's magnetostriction measurements\citep{VectorPRL}. According to their capacitance dilatometer measurements, at approximately 25 mK, for B//$ab$-plane and original sample length $L=0.9144$ mm, the dilation change  $\Delta L_{sample}$ at the transition point $H_c$ is $\sim 5$ nm.  The voltage signal $\Delta V$ at the same transition point is $\sim 0.7$ $\mu$V.
So the calibration relation can then be estimated as
\begin{equation} \Delta V/ \Delta L\approx 140 \mu V/\mu m
\label{eq2}
\end{equation} 

This estimation agrees with the room temperature value found in the reference\citep{Takahashi2002} within a factor of 2 (PRC400, with $V_{bias}$=0.05 V). In fact, because the different cantilevers have different ``cell effect", the signal at the transition point is also affected. We performed measurements on the same $CeCoIn_5$ sample using different cantilevers. At the transition point, the largest $\Delta V$ signal could be up to twice large as the smallest one. The $\Delta V$ shown above is an average value. Therefore, this calibration relation can be used for some rough estimations, but cannot be used to determine absolute sample length changes.

With equation \ref{eq2}, we can estimate our low temperature measurement resolution. As previously mentioned, the lowest peak to peak noise level is $\sim$ 20 nV, so the best resolution we can achieve is $\sim$2 $\AA$. With this calibration, the temperature and field dependence of the cell effect also can be roughly estimated to be $\sim$3 nm/Kelvin and $\sim$1.5 nm/Tesla respectively. These are average values obtained by measuring 3 different cantilevers. Note these estimated values only apply for low temperature measurements in the millikelvin range. This cell effect includes the result of resistor mismatch, piezoresistor magnetoresistance, the expansion of silicon, etc. But in fact, for many low temperature measurements, one is interested in measuring a phase transition or seeing quantum oscillations, and this cell effect could be subtracted as the background signal. So despite the "cell effect", this technique is still very useful for detecting dilation changes caused by phase transitions for samples with limited size. Moreover, for samples that are large enough, a very useful simultaneous multi-axis measurement can be performed with the setup shown in Fig.\ref{1} (a). 

\section{Phase transition measurement}
\begin{figure}
\includegraphics[width=0.9\linewidth]{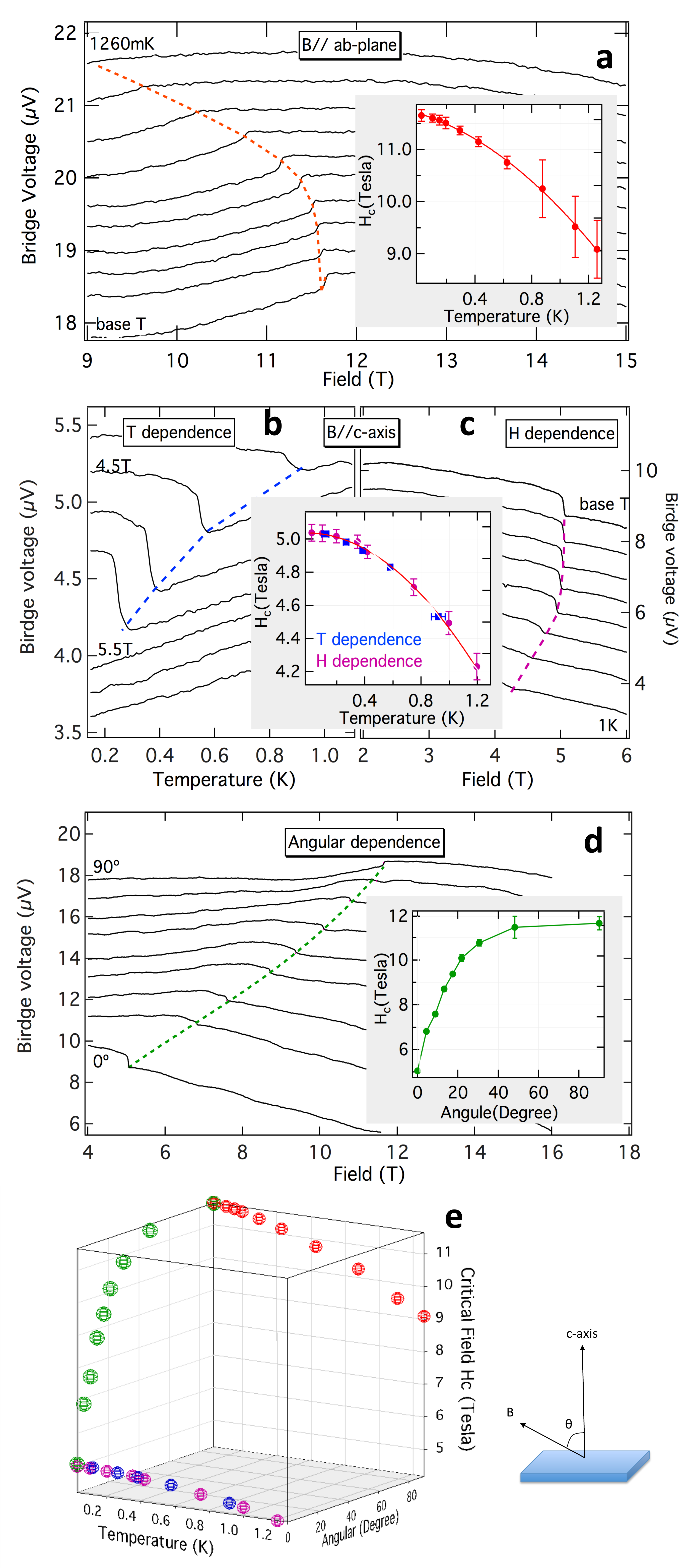}% % Important NOTE: Please make certain your figures do not include local directory paths. ex."c:\file\sub\fig1.eps" 
\caption{Graph (a) shows the field dependence results of
$CeCoIn_5$ at different temperatures (top to bottom curves are taken from 1.26 K to base temperature). The field is applied in the $ab$-plane ($\theta$ = 90$^{\circ}$). Insert graph is the phase diagram. Graphs (b) and (c) show the results of the field applied to $c$-axis ($\theta$ = 0$^{\circ}$). Graph (b) shows the temperature dependence at different fields (top to bottom data are taken at field from 4.5 T to 6 T). Graph (c) shows the field dependence at different temperatures (top to bottom data are taken from base temperature to $\sim$1 K). The phase diagram based on these two graphs is placed in the middle. Graph (d) shows the field dependence at the different angle (top to bottom data are taken from $\theta$ = 90$^{\circ}$ to 0$^{\circ}$). The bottom graph (e) is the 3D view of a summary of all three phase diagrams.}% 
\label{CeCoIn5} 
\end{figure}

The heavy-fermion superconductor 115 family, $REMIn_5$ ($RE = La$ or $Ce$; $M$ = $Co, Rh$ or $Ir$), has been extensively investigated in the last decade owing to several unusual properties of their superconducting (SC) states\citep{0953-8984-13-17-103, PhysRevLett.86.5152}. The tetragonal crystal structure alternates magnetic $REIn_3$ and non-magnetic $MIn_2$ layers along the $c$-axis. One of the most interesting members of this family is $CeCoIn_5$, not only because it demonstrates a sharp, clear first-order phase transition from the superconducting state to the normal state at high magnetic fields, but also because it is the first material to exhibit a Fulde-Ferrell-Larkin-Ovchinnikov (FFLO) superconducting state\citep{PhysRevLett.91.187004,VectorPRL}.

The $CeCoIn_5$ sample we used for this measurement is 1$\times$1$\times$1.5 (a$\times$b$\times$c) mm$^3$ in size. The sketch at the right bottom of Fig.\ref{CeCoIn5} shows the orientation between the sample and applied magnetic field. The cantilever tip gently rested on top of the sample and the signal shown here reflected only the changes of the $c$-axis. Here $\theta$ is defined as the angle between the magnetic field and the $c$-axis.

As shown in Fig.\ref{CeCoIn5}, TE and MS measurements were made at different directions and temperatures, which provided a complete 3D-phase diagram. Fig.\ref{CeCoIn5} (a) shows the field dependence results of $CeCoIn_5$ at different temperatures. From top to bottom, the data were taken at 1260 mK, 1100 mK, 900 mK, 600 mK, 450 mK, 300 mK, 200 mK, 150 mK, 100 mK and 25 mK respectively. For this graph, a magnetic field was applied along the $ab$-plane ($\theta$ = 90$^{\circ}$). The inset to graph (a) is the resulting phase diagram. Fig.\ref{CeCoIn5} (b) and (c) are the results of the field applied to $c$-axis ($\theta$ = 0$^{\circ}$). Fig.\ref{CeCoIn5} (b) is the temperature dependence at different fields (4.5 T, 4.8 T, 4.9 T, 4.95 T, 5 T, 5.5 T, 6 T), while graph (c) shows the field dependence at different temperatures (25 mK, 100 mK, 200 mK, 380 mK, 420 mK, 750 mK, 1000 mK, 1200 mK). The resulting phase diagram is shown in the inset of Fig.\ref{CeCoIn5} (b) and (c). Graph (d) shows the field dependence at different angles ($\theta$ = 0$^{\circ}$, 4$^{\circ}$, 9$^{\circ}$, 13$^{\circ}$, 17$^{\circ}$, 22$^{\circ}$, 30$^{\circ}$, 50$^{\circ}$, 90$^{\circ}$ respectively).

These overall magnetic field versus temperature phase diagrams are in excellent agreement
with previous work\citep{VectorPRL,PhysRevB.71.020503,PhysRevB.71.020503,PhysRevB.70.134513,PhysRevB.70.020506}.  Most importantly, a complete 3D phase diagram can be made. (See the last graph (e) in Fig.\ref{CeCoIn5}). By decreasing the temperature from 1.2 K to the base temperature ($\sim$25 mK), the transition point $H_c$ increased from $\sim$9 T to 11.65 T. While at base temperature, the sample was rotated 90 degrees from the B//$ab$-plane position to the B//$c$-axis position. The transition critical field $H_c$ decreased from 11.65 T to $\sim$4.98 T (see the $\theta$-$H$ plane at $T = 25$ mK). Finally, at position B//$c$-axis, the temperature was increased from 25mK to 1.2 K. The transition point continued to decrease and finally vanished around 4.1 T (see the $H$-$T$ plane at $\theta$ = 0$^{\circ}$).

\section{Simultaneous Measurement of two sample axes}
\begin{figure} 
\includegraphics[width=1\linewidth]{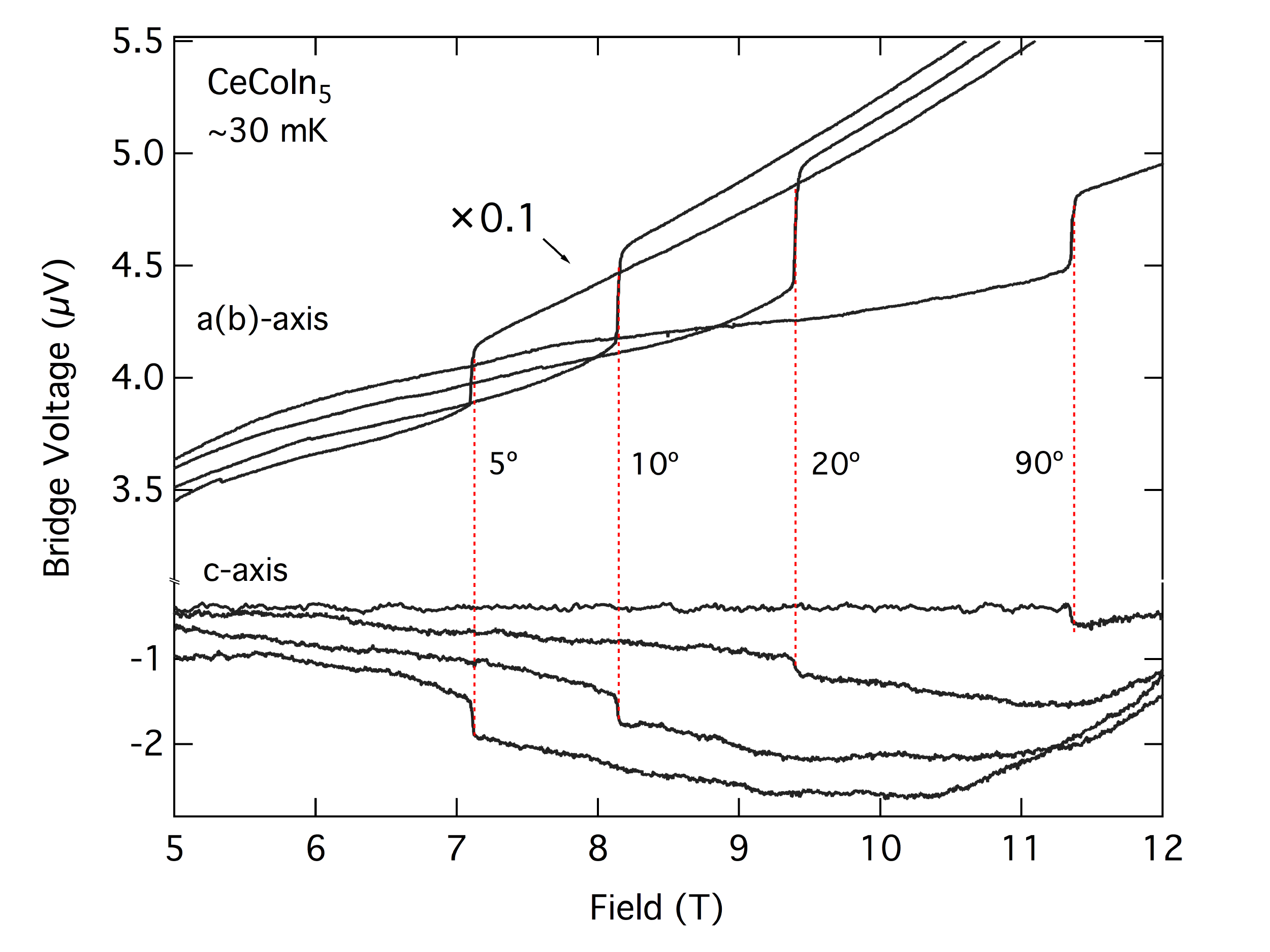}% % Important NOTE: Please make certain your figures do not include local directory paths. ex."c:\file\sub\fig1.eps" 
\caption{\label{2axisCeCoIn} Angular dependence data for both $c$ and $a$($b$)- axis taken at the same time. Note that the data collected along the $a$($b$)-axis is decreased by a factor of 10 to be shown on a similar scale.}% 
\end{figure}

The same $CeCoIn_5$ sample used in the previous section was also used for these measurements. The orientation between the sample and applied magnetic field is shown in the lower right of Fig.\ref{CeCoIn5}. This time two cantilevers were used to measure the two different axis of the sample simultaneously.  As can be seen in Fig.\ref{1} (c), $c$-axis is perpendicular to the sample surface which is directly facing out of the page (in Fig.\ref{1} (a), the surface faces up), and $a$($b$)-axis is perpendicular to the side surface (in Fig.\ref{1} (a), the surface faces aside). The signals will reflect the changes of $c$ and $a$($b$)-axis at the same time.

Fig.\ref{2axisCeCoIn} shows two groups of angular dependent data, one group (upper in the graph) is the measurement along the a(b)-axis and the other is the measurement along the $c$-axis (lower part in the graph). Both groups contains four curves that are taken at $\theta$ = 5$^{\circ}$, 10$^{\circ}$, 20$^{\circ}$, 90$^{\circ}$ respectively. As can be seen, the transitions for these two groups of data correspond to each other. At the transition point, the signal, $\Delta V$ for the $a$($b$)-axis is more than 10 times larger than the signal for $c$-axis. For clarity, the $y$-axis of the graph was split and data for $a$($b$)-axis are decreased by a factor of 10 in order to get a similar scale with the data for $c$-axis. For the measurements performed on the $c$-axis, even though a different cantilever was used than in section III, the signal amplitude for both measurements are still comparable, despite the different background caused by the cell effect. Thus for the measurements along the $a$($b$)-axis, the large $\Delta V$ at the transition point indicated the dilation change along $a$($b$)-axis is much bigger than that along $c$-axis given the crystal dimensions. This is very important information which can help to estimate the volume change and understand the whole lattice.

\section{Measurement of quantum oscillations} 
\begin{figure}
\includegraphics[width=1\linewidth]{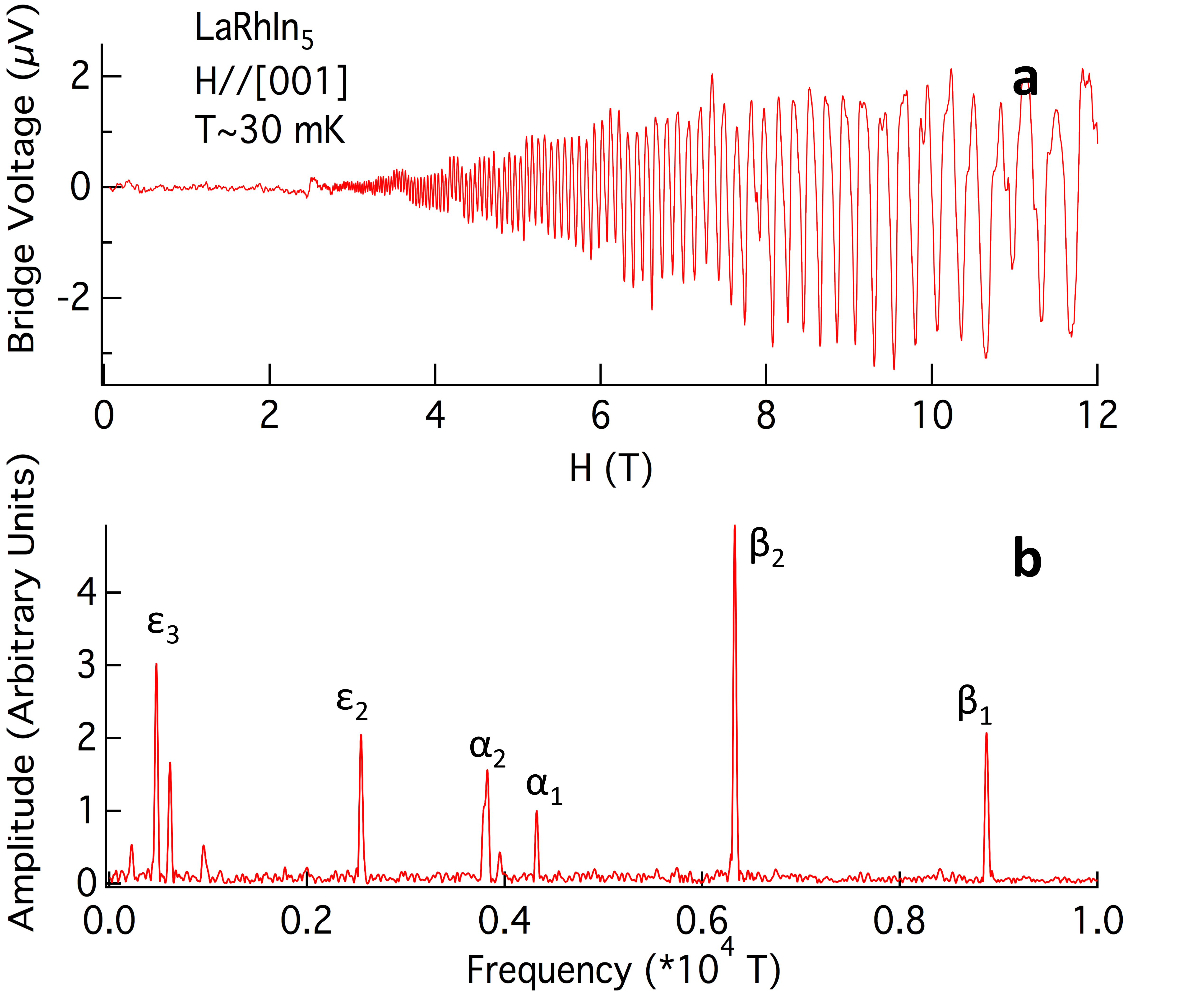}% % Important NOTE: Please make certain your figures do not include local directory paths. ex."c:\file\sub\fig1.eps" 
\caption{\label{LaRhIn5-1}(a) Background subtracted trace showing quantum oscillations and (b) the resulting FFT spectrum in $LaRhIn_5$}% 
\end{figure}

In addition to the application for simultaneous multi-axis measurement, the advantages of our cantilever dilatometer manifest in other aspects. The use of cantilevers in pulsed magnets and the oscillatory magnetic torque measurements were successfully studied\citep{PhysRevB.80.241101, adhikari:013903, PhysRevB.81.184506}. The dilation measurements on $LaRhIn_5$ proved the cantilever can detect quantum oscillations via magnetostricton and the signal is significant.

The $LaRhIn_5$ sample we used here is much thinner, only 1$\times$1$\times$0.1 mm$^3$(a$\times$b$\times$c) in size. The sample and field position are similar to the $CeCoIn_5$ sample shown in Fig.\ref{CeCoIn5} bottom. The cantilever also touches the top of the sample, which measures the dilation change of $c$-axis. 

\begin{figure} 
\includegraphics[width=1\linewidth]{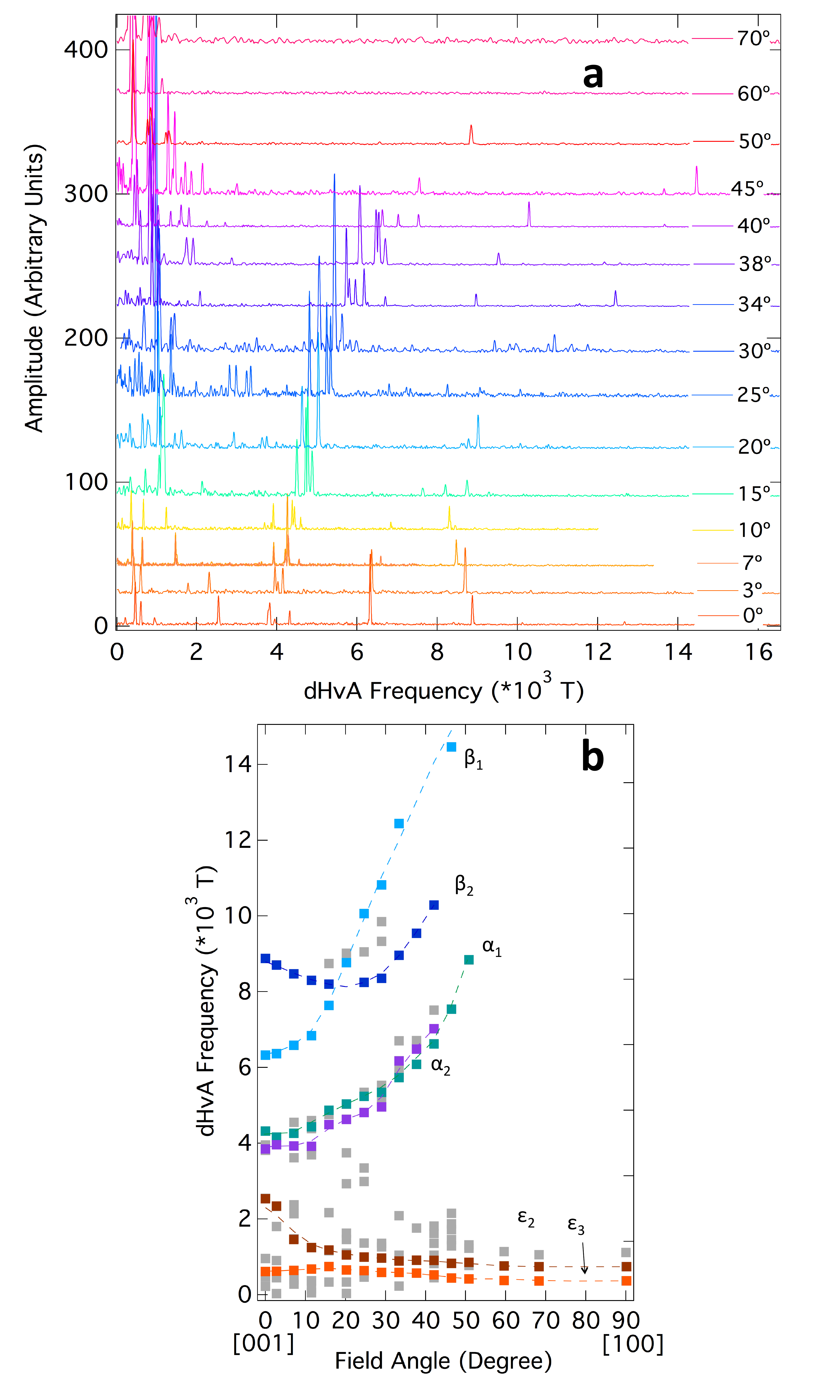}% % Important NOTE: Please make certain your figures do not include local directory paths. ex."c:\file\sub\fig1.eps" 
\caption{\label{LaRhIn5-2}Angular dependence of the frequency in $LaRhIn_5$. (a)FFT  spectra at different angle and (b) selected peaks vs. angle. }% 
\end{figure}

In Fig.\ref{LaRhIn5-1}, data from $LaRhIn_5$ with the field parallel to the $c$-axis is shown. In the lower part of the graph, a fast Fourier transform (FFT) spectrum for the $LaRhIn_5$ analysis for fields between 4 T and 12 T is shown. The upper graph shows the clear oscillation data up to 12 T. The quality of this data is typical for all of the investigated trace curves in Fig.\ref{LaRhIn5-2}. Multiple frequencies are found in the FFT spectrum; main fundamental branches are $\alpha_1$, $\alpha_2$, $\beta_1$, $\beta_2$, $\varepsilon_1$ and $\varepsilon_2$.

In addition to high resolution, this cantilever dilatometer can be easily rotated in the cryostat. For quantum oscillation measurements, the data taken at different angles will reflect the Fermi surface at that direction, so obtaining the angular dependent data is important for understanding the full Fermi surface. Because of the confined space and liquid/gas environment at field center, sample rotation in other dilation measurements is rather limited. However, this piezo cantilever dilatometer does not have these restrictions, making it the perfect choice to perform angular dependent measurements at low temperature and high field.

Figures \ref{LaRhIn5-2} (a) and (b) show the angular dependence of the oscillation frequencies of $LaRhIn_5$, both in the FFT spectra and selected peaks from those spectra. The branches $\alpha_i$ ($i$=1, 2) and $\beta_2$ roughly follow the 1/$cos\theta$ dependence, where $\theta$ indicates a field angle tilted from [001] to [100]. In comparison to the theoretical calculation in Shishido's paper\citep{JPSJ.71.162}, this angular dependence suggests that the corresponding Fermi surface is nearly cylindrical. Because of the theoretical calculation, branch $\alpha_2$ is due to a band 15-electron Fermi surface, which is nearly cylindrical, but corrugated, allowing maximum and minimum cross-sections. Branch $\beta_i$ is due to a highly corrugated band 14-electron Fermi surface. A feature of $\beta_i$ is a minimum in its frequency at $\sim$30$^{\circ}$ from the $c$-axis. This feature is shown in our data in Fig.\ref{LaRhIn5-2} (b). Branch $\varepsilon_i$ is due to a band 13-hole Fermi surface whose topology is similar to a lattice. In previous studies\citep{JPSJ.70.2248,Onuki200213,PhysRevB.79.033106}, these low frequency signals have never been clearly shown. However, by using the cantilever dilatometer, the low frequency features are well resolved in the oscillatory magnetostriction measurements.

The angle-dependent quantum oscillations in $LaRhIn_5$ at $\sim$25 mK for a wide angle region have been successfully observed and shown in Fig.\ref{LaRhIn5-2}. It is clearly found that these main frequencies which are corresponding to certain Fermi surfaces, systematically shifted with rotation angle $\theta$. The significant signals at low frequency range may need further investigation.

\section{Conclusion}

Advancing our previous work, we improved the design implementation of the miniature size AFM cantilever based
dilatometer. The successful applications of our new design on heavy Fermion SC samples $REMIn_5$ show that the cantilever dilatometer is a very useful tool for magnetoelastic and quantum oscillation investigations. Compared to the
traditional capacitance dilatometer and FBGs, although this dilatometer can not provide accurate length changes of a sample, it still has unique merits. First, the sample used for this dilatometer can be very small (smaller than 0.1 mm in length). This is already beyond the limitations of many other dilatometer forms. Also, the rotation measurement is usually considered to be a big challenge for capacitance and FBGs dilatometers. Since the volume of the piezo dilatometer can be rather small, it can be easily rotated in very constrained spaces, like field center in a dilution refrigerator, without gravity effects, providing another advantage for this technique. Its versatility makes it suitable for measurements under the many conditions imposed by low temperatures (like liquid or gaseous helium) and high magnetic fields. Moreover, this technique shows suitability of application in oscillatory magnetostricton measurements, measuring even high frequency signals easily. The final advantage of our dilatometer is shown by the successful application on simultaneous multi-axis  dilation measurements. The dilation measurement for more than one direction at the same time is difficult for other dilatometer techniques, which makes our cantilever dilatometer a unique tool for this type of measurement.

\begin{acknowledgements} We would like to thank Vaughan Williams for mechanical
construction work and James Andrew Powell for electrical design and construction. Support for
this work was provided by the DOE/NNSA under Grant No. DEFG52-06NA26193. Work at Occidental College was supported by the National Science Foundation under DMR-1006118. This
work was performed at the National High Magnetic Field Laboratory, which is
supported by National Science Foundation Cooperative Agreement No. DMR-0654118,
the State of Florida, and the U.S. Department of Energy. \end{acknowledgements}

\end{document}